\documentclass[aip,jcp,reprint]{revtex4-1}
\usepackage{dcolumn}
\usepackage{bm}
\usepackage{amsfonts}
\usepackage{amsthm}

\begin{document}

\title{Notes on configurational thermostat schemes}

\author{A.A. Samoletov}
\email{samolet@fti.dn.ua}
\affiliation{Institute for Physics and Technology, NASU, 
83114 Donetsk, Ukraine}
\author{C.P. Dettmann}
\email{Carl.Dettmann@bris.ac.uk}
\affiliation{Department of Mathematics, University of Bristol, 
Bristol BS8 1TW, United Kingdom}
\author{M.A.J. Chaplain}
\email{chaplain@maths.dundee.ac.uk}
\affiliation{Division of Mathematics, University of Dundee, Dundee DD1 4HN,
United Kingdom}
%
\maketitle

\section{Introduction}

The simulation of rheological properties of molecular fluids requires realistic
temperature control correctly taking into account the streaming velocity of the
fluid.  This has become feasible following the introduction of thermostatting
techniques defining temperature using only configurational variables. 
Configurational approaches are also important in many other contexts, for
example in systems subject to an external pressure constraint and in
understanding conformational dynamics of biomolecules.

In a series of papers \citep{Braga2005,Braga2006,travis2006,travis2008}
Braga and Travis proposed a Nos\'e-Hoover type thermostat based on a
configurational definition of temperature (in short BT thermostat), removing
difficulties associated with the earlier Delhomelle-Evans approach
\citep{Delhommelle2001} which
for example does not preserve the canonical distribution in the equilibrium
case. Recently the BT thermostat has had further attention,
development and application in the literature
\citep{Jepps2010,hoover2009,hoover2009a,hunt2009,Desgranges2008,McWhirter2008}.
In the principal paper \citep{Braga2005}, the authors considered the
configurational
thermostat scheme proposed in references \citep{Samoletov2004a,SDC07} (in
short SDC scheme) as a starting point
and then presented their own configurational temperature thermostat
as a further development of the SDC scheme. Correct action of both 
BT and SDC thermostats has been confirmed by a variety of numerical
simulations in these works.

In this note, we link together two above-mentioned contributions in the
development of configurational thermostats and thus accomplish two goals.
First, we show that the configurational
temperature BT thermostat is a particular case of the SDC scheme. 
Second, we demonstrate how this particular case benefits
from the general method \citep{Samoletov2004a,SDC07} in a variety
of aspects including the physical context, a stochastic counterpart thermostat
(to the authors' knowledge this has not previously appeared in the
literature) and ergodicity issues. 

\section{The configurational temperature thermostat}

Let us clearly state in what way the BT thermostat
is a particular case of the SDC scheme.
The general SDC scheme \citep{Samoletov2004a,SDC07}
involves two different definitions of the configurational temperature
and the corresponding double temperature control (for example, to allow
control at two timescales), coupling between thermostats, anisotropy, 
and a series of definitions of the thermostat variable that does not
directly relate to a particular temperature control. To reduce this
general configurational thermostat method to the BT 
equations (our first goal), consider the simplified version of the SDC 
where the virial temperature control
and the coupling of thermostat variables are omitted:%
\begin{eqnarray*}
m_{k}\mathbf{\dot{q}}_{k} & = &
-\tau\bm{\nabla}_{\mathbf{q}_{k}}V(\bm{q})+\bm{\xi}_{k},\\
\dot{\tau} & = &
\frac{1}{Q_{\tau\tau}}\sum_{k=1}^{N}\frac{1}{m_{k}}\left[(\bm{\nabla}_{\mathbf{q
}_{k}}V)^{2}-k_{\mathrm{B}}T\Delta_{\mathbf{q}_{k}}V\right],
\end{eqnarray*}
Then consider the dynamic variables $\boldsymbol{\xi}_{k}$ according
to the case $(a)$ in references \citep{Samoletov2004a,SDC07} (variables
$\boldsymbol{\xi}$
for all particles of the system are varied independently),
\[
\bm{\dot{\xi}}_{k}=-\frac{1}{Q_{\xi\xi}}\frac{1}{m_{k}}\bm{\nabla}_{\mathbf{q}_{
k}}V(q),
\]
set $Q_{\xi\xi}=1$, and finally define the canonical variables,
$\mathbf{r}_{k}=\sqrt{m_{k}}\mathbf{q}_{k},\quad\mathbf{p}_{k}=\sqrt{m_{k}}
\boldsymbol{\xi}_{k}.$
As a result, we arrive at the BT equations as a particular
case of the more general SDC scheme, 
\begin{eqnarray}
\mathbf{\dot{r}}_{k} & = &
\frac{\mathbf{p}_{k}}{m_{k}}-\tau\bm{\nabla}_{\mathbf{r}_{k}}V,\quad\mathbf{\dot
{p}}_{k}=-\bm{\nabla}_{\mathbf{r}_{k}}V,\nonumber \\
\dot{\tau} & = &
\frac{1}{Q_{\tau\tau}}\sum_{k=1}^{N}\left[(\bm{\nabla}_{\mathbf{r}_{k}}V)^{2}-k_
{\mathrm{B}}T\Delta_{\mathbf{r}_{k}}V\right].\label{eq:TB-SD}
\end{eqnarray}
It is easy to show that the canonical distribution, 
$\exp[-\frac{1}{k_{\mathrm{B}}T}(\sum\frac{\mathbf{p}^{2}}{2m}+V)]$,
is invariant for this dynamics.

In the following paragraphs we demonstrate in what way the SDC scheme 
contributes to the understanding of the configurational temperature thermostats. 

A frequent problem, even if the canonical measure is invariant for
the dynamics (\ref{eq:TB-SD}), is that only ergodic dynamics correctly samples
the canonical distribution. In general, this condition is very difficult
to prove for systems with many degrees of freedom. In particular, simple
one-dimensional systems \citep{Legoll2009}
reveal a difficulty and provide an important challenge for deterministic
thermostatting methods. On the other hand, stochastic dynamics
is typically ergodic \citep{Leimkuhler2009,Leimkuhler2010}.

\section{The stochastic counterpart}

The Nose-Hoover thermostat is a deterministic counterpart of the Langevin
stochastic dynamics (\textit{e.g. \citep{SDC07}}). This fact is useful
to determine the appropriate characteristic time scale of the dynamics.
To clarify the physical sense of the configurational temperature thermostat,
we formulate a stochastic counterpart of the dynamics (\ref{eq:TB-SD}).
For the sake of brevity we provide only the stochastic
dynamics and its justification. A formal analysis 
can be done following the method presented in references 
\citep{Samoletov1999,Mazo2002}.

A particle of mass $m$ in a dissipative medium under the influence of
a constant force $F$, has achieved the constant velocity $v_{s}=\tau F$,
where $\tau$ is the mobility, after the transient time, $\sim m\tau$.
We take into account the mobility as a perturbation of velocities.
Explicitly, consider the following stochastic equations of motion:
\begin{eqnarray}
\mathbf{\dot{q}}_{k}&=&\frac{\mathbf{p}_{k}}{m_{k}}-\tau\boldsymbol{\nabla}_{
\mathbf{q}_{k}}V(q)+\sqrt{2D}\mathbf{f}_{k}(t),\nonumber \\
\dot{\mathbf{p}}_{k}&=&-\boldsymbol{\nabla}_{\mathbf{q}_{k}}V(q),\label{eq:ISDE}
\end{eqnarray}
where $\{\mathbf{f}_{k}(t)\}$ is the set of independent standard
``white noises''. The Fokker-Planck equation corresponding to
(\ref{eq:ISDE}) has the form,
\begin{eqnarray}
\partial_{t}\rho=\sum_{(k)}\{&-&\boldsymbol{\nabla}_{\mathbf{q}_{k}}[(\frac{
\mathbf{p}_{k}}{m_{k}}-\tau\boldsymbol{\nabla}_{\mathbf{q}_{k}}V)\rho] \nonumber
\\
&+&\boldsymbol{\nabla}_{\mathbf{p}_{k}}\cdot(\boldsymbol{\nabla}_{\mathbf{q}_{k}
}V\rho) 
+D\triangle_{\mathbf{q}_{k}}\rho\}\equiv\mathcal{F}^{*}\rho, \label{eq:IFP}
\end{eqnarray}
where $\mathcal{F}^{*}$ is the Fokker-Planck operator.

These stochastic dynamics can be generalized to the case of the state
dependent mobility, $\tau\rightarrow\tau\varphi(p)$, where $\varphi(p)$
is a positive function of momenta.
The Fokker-Planck equation $\partial_t\rho=\mathcal{F}_{\varphi}^{*}\rho$
corresponding to that case  
has the form (\ref{eq:IFP}) where
$ \tau\rightarrow\tau\varphi(p)$ and $D\rightarrow D\varphi(p)$.
Indeed, suppose 
$\tau$ and $D$ are connected to each other by the relation
$D=\tau k_{B}T$.  
Substituting the canonical distribution,
$\rho\sim\exp[-\beta(\sum_{(k)}(2m_{k})^{-1}\mathbf{p}_{k}^{2}+V(q))]$,
into $\mathcal{F}_{\varphi}^{*}\rho$ 
and  then applying the condition, $D=\tau k_{B}T$, we 
arrive at the identity,
$\mathcal{F}_{\varphi}^{*}\rho\equiv0$.
Thus the canonical measure,
$d\mu\sim\rho\prod_{(k)}d\mathbf{q}_{k}d\mathbf{p}_{k}$
 is invariant for this dynamics  
and we obtain the stochastic
counterpart of the deterministic configurational thermostat (\ref{eq:TB-SD}).
The standard case
is $\varphi(p)\equiv1$, $\mathcal{F}^{*}\rho\equiv0$. The more general case, 
$\varphi=\varphi(p,q)$, can be considered but in that case we must specify
the interpretation of the stochastic equations (It\^o or Stratonovich)
and correspondingly add an extra drift term into (\ref{eq:ISDE}).

\section{The extended configurational thermostat}

Following the general SDC scheme \citep{Samoletov2004a,SDC07}
the deterministic BT 
thermostat (\ref{eq:TB-SD})
can be refined in the following manner. Consider the extended dynamical
system:
\begin{eqnarray}
\mathbf{\dot{r}}_{k} & = &
\frac{\mathbf{p}_{k}}{m_{k}}-\tau\bm{\nabla}_{\mathbf{r}_{k}}V+\boldsymbol{\xi}_
{k},\quad\mathbf{\dot{p}}_{k}=-\bm{\nabla}_{\mathbf{r}_{k}}V,\nonumber \\
\dot{\tau} & = &
\frac{1}{Q_{\tau\tau}}\sum_{k=1}^{N}\left[(\bm{\nabla}_{\mathbf{r}_{k}}V)^{2}-k_
{\mathrm{B}}T\Delta_{\mathbf{r}_{k}}V\right],\quad\boldsymbol{\dot{\xi}}_{k}
=\mathbf{h}_{k,}\label{eq:TB2SD}
\end{eqnarray}
 where $\{\boldsymbol{\xi}_{k}\}$ are additional dynamical variables.
Functions $\{\mathbf{h}_{k}\}$ can be fixed in a variety of forms
as described in references \citep{Samoletov2004a,SDC07}. For example, 
set
$\mathbf{h}_{k}=-\frac{1}{Q_{\xi}}\sum_{(i)}\boldsymbol{\nabla}_{\mathbf{r}_{
i}} V$
for all particles of the system. Then the canonical measure (augmented
with the Gaussian measure for thermostat variables $\tau$ and
$\boldsymbol{\xi}$)
is invariant for dynamics (\ref{eq:TB2SD}). 
In that case, the velocity of the center of inertia of a system dynamically
fluctuates
around zero with mobility $Q_{\xi}^{-1}$, where $Q_{\xi}$ is a parameter.

\section{On ergodicity}

The thermostat dynamics is able to correctly sample
the canonical distribution only if the motion is ergodic. To improve
ergodicity, we proposed in reference \citep{SDC07} to perturb the dynamics
of the augmented thermostat variables (the Gaussian invariant measure
for these variables is supposed) by additional random noise. This
method was further investigated together with mathematical justification
in references \citep{Leimkuhler2009,Leimkuhler2010}. Numerical effectiveness of
this method was also confirmed  
\citep{SDC07,Leimkuhler2009,Leimkuhler2010}. Taking into account
additional advantages of this method including the ``gentle'' action of the
stochastic term \citep{Leimkuhler2009,Leimkuhler2010},
we propose to apply it to configurational temperature thermostats
(\ref{eq:TB-SD}), (\ref{eq:TB2SD}) similarly to the
description in reference \citep{SDC07}.  For example, in set (\ref{eq:TB-SD})
we propose to perturb the dynamics of 
$\tau$-variable by the ``white noise'' $f(t)$,%
\[
\dot{\tau}=\frac{1}{Q_{\tau\tau}}\sum_{k=1}^{N}\left[(\bm{\nabla}_{\mathbf{r}_{k
}}V)^{2}-k_{\mathrm{B}}T\Delta_{\mathbf{r}_{k}}V\right]-\Lambda\tau+\sqrt{
2\mathrm{D}}f(t),
\]
where (positive) $\Lambda$ and $\mathrm{D}$ are connected to each
other by the relation, $k_{\mathrm{B}}T\Lambda=\mathrm{D}Q_{\tau\tau}$. Then the
augmented canonical measure is invariant for the dynamics.
We conjecture that this is the unique 
invariant measure (compare with\citep{Leimkuhler2009}).

\section{Conclusion}

In this Note, we have linked together two contributions in the
development of configurational thermostats, the BT thermostat and
the SDC scheme. 
We have shown that proposed configurational 
thermostats \citep{Braga2005,Braga2006,travis2006,travis2008}
are generalized and enriched in both understanding and content
by the SDC scheme. 
We have presented the stochastic counterpart (\ref{eq:ISDE}) to the
configurational thermostat (\ref{eq:TB-SD}). 
%
\end{document}